# Real-Space Visualization of Site-Selective Co 3*d* States in the Ising Magnet $Ca_3Co_2O_6$

Kamini Gautam[1,†], Shunsuke Kitou[1,2,†,*], Yuiga Nakamura[3], Norimasa Sasabe[4], Arvind Kumar Yogi[5], Dinesh Kumar Shukla[5], Yuichi Yamasaki[1,4,6], and Taka-hisa Arima[1,2,‡]

[1]*RIKEN Center for Emergent Matter Science (CEMS), Wako 351-0198, Japan*
[2]*Department of Advanced Materials Science, The University of Tokyo, Kashiwa 277-8561, Japan*
[3]*Japan Synchrotron Radiation Research Institute (JASRI), SPring-8, Hyogo 679-5198, Japan*
[4]*Center for Basic Research on Materials (CBRM), National Institute for Materials Science (NIMS), Tsukuba, 305-0047, Japan*
[5]*UGC-DAE Consortium for Scientific Research, Khandwa Road, Indore 452001, India*
[6]*International Center for Synchrotron Radiation Innovation Smart, Tohoku University, Sendai 980-8577, Japan*

High-spin $Co^{3+}$ ions can retain a finite orbital contribution and exhibit pronounced magnetic anisotropy, providing model systems for Ising magnetism. However, direct real-space observations of their orbital states remain limited. Here, we visualize the site-dependent Co 3*d* electronic states in the Ising-chain magnet $Ca_3Co_2O_6$, which contains crystallographically inequivalent high-spin and low-spin $Co^{3+}$ sites, using synchrotron X-ray diffraction and valence electron density (VED) analysis based on core differential Fourier synthesis. The octahedral Co site exhibits a VED distribution consistent with a low-spin state, whereas the trigonal-prismatic Co site shows pronounced anisotropy that cannot be reproduced by crystal electric field (CEF) effects alone. A wavefunction analysis incorporating CEF effects, spin–orbit coupling, and ligand-mediated Co 3*d*– O 2*p*–Co 4*p* hybridization reproduces the observed VED anisotropy. These results establish a site-selective real-space picture of the Co electronic states and demonstrate the capability of VED analysis to probe local orbital states and orbital hybridization in quantum materials.

## 1. Introduction

The crystal electric field (CEF) and spin–orbit coupling (SOC) play crucial roles in determining the orbital states and magnetic anisotropy of transition-metal compounds[1)-2)]. Together with electron correlations and orbital-dependent hopping, they govern the form and anisotropy of magnetic interactions. For instance, the spin–orbital entangled $J_{\mathrm{eff}}$ = 1/2 state is a key feature for realizing the Kitaev spin liquid[3)]. In 3$d$ transition metal compounds, the CEF generally lifts the degeneracy of the $d$ orbitals, suppressing the orbital degrees of freedom and quenching orbital angular momentum. However, in high-spin Co ions, SOC often partially restores orbital angular momentum, leading to characteristic magnetic anisotropy. Ising-chain compounds such as $CoNb_2O_6$ and $Ca_3Co_2O_6$ therefore provide important platforms for studying low-dimensional and frustrated magnetism[4)-6)].

$Ca_3Co_2O_6$ is a canonical frustrated Ising-chain magnet crystallizing in the rhombohedral space group $R\bar{3}c$ [Fig. 1(a)][6),7)]. The quasi-one-dimensional Co chains are arranged to form a triangular lattice. The intrachain and interchain magnetic interactions are predominantly ferromagnetic and antiferromagnetic, respectively. The geometrically frustrated triangular lattice contributes to its unusual magnetic and dielectric properties[6)-15)], alongside its Ising-like character. The magnetic behavior of $Ca_3Co_2O_6$ is influenced by CEF splitting, Hund's coupling, and SOC. This material contains two types of $Co^{3+}$ ions, with distinct site symmetries and spin states [Fig. 1(a)]. Co1 and Co2 ions, arranged alternately along the $c$-axis, are coordinated by six oxygen atoms, forming trigonal-prismatic and octahedral coordination environments [Fig. 1(b)], respectively. Previous studies have established that Co1 ions are in the high-spin state ($S$ = 2), while Co2 ions are in the low-spin state ($S$ = 0)[12)-21)].

The origin of Ising character in $Ca_3Co_2O_6$ has been widely debated. Soft X-ray spectroscopies, including absorption and magnetic circular dichroism, probe the valence, spin, and orbital states of Co ions[18)], revealing a considerable orbital magnetic moment along the $c$-axis and strong magnetocrystalline anisotropy. The contributions of CEF splitting, Hund's coupling, and SOC to the orbital and spin states in $Ca_3Co_2O_6$ are supported by theoretical studies[10),16),19),21)]. More recently, $s$-core-level non-resonant inelastic X-ray scattering ($s$-NIXS) has revealed a complex orbital state of the minority-spin electron at the high-spin Co1 site[20)], providing important insights into the microscopic origin of the anisotropy. However, despite these advancements, the detailed site-dependent Co-3*d* electronic states remain unresolved. Notably, the coexistence of distinct spin and orbital states at two inequivalent Co sites complicates their site-resolved characterization. Although X-ray photoemission spectroscopy[17)], X-ray absorption[18)], and *s*-NIXS[20)] provide detailed information on the electronic states, they do not directly reconstruct the local electron-density distributions independently at the two Co sites.

In this study, we apply core differential Fourier synthesis (CDFS) analysis[22),23)], based on synchrotron X-ray diffraction, to directly visualize the valence electron density (VED) distribution for the two inequivalent Co sites independently in $Ca_3Co_2O_6$. This technique enables the visualization of the entire VED distribution within the unit cell by performing Fourier synthesis using structure factors derived from the measured diffraction intensities in various systems[22)-28)]. The reconstructed VED provides quantitative insights into the roles of CEF effects, SOC, and complex orbital hybridization. This study paves the way for a deeper understanding of the intricate interplay between CEF splitting and SOC in quantum materials.

## 2. Methods

A rod-like single crystal of $Ca_3Co_2O_6$ was grown using the flux technique. Initially, a polycrystalline sample was prepared using the standard solid-state approach[29)]. Stoichiometric amounts of $CaCO_3$ and $Co_3O_4$ powders were mixed and then ground together for 12 h. The mixture was subsequently heated in air at 900 °C for 24 h. After intermediate grinding, the powder was pelletized and sintered at 1000°C for 48 h, followed by cooling at a rate of 5 °C/min down to room temperature. The resulting $Ca_3Co_2O_6$ compound was then combined with $K_2CO_3$ flux in a 1:10 weight ratio and slowly heated at a rate of ~ 0.6 °C/min up to 950°C. The temperature was kept for 24 h before slow cooling at a rate of 2 °C/h to 700°C and then further cooling at a rate of 5 °C/h to room temperature. As a result, rod-shaped crystals approximately 5 mm in length were grown, with the *c*-axis running parallel to the rod axis.

X-ray diffraction experiments were performed on BL02B1[30)] at the SPring-8 synchrotron facility, Japan. An $N_2$-gas-flow cryostat was employed to cool the crystals to 100 K. A two-dimensional CdTe PILATUS detector, which had a dynamic range exceeding $10^6$, was used to record the diffraction pattern. The X-ray wavelength was calibrated to be λ = 0.30918 Å. The intensities of Bragg reflections of the interplanar spacing $d > 0.28$ Å were integrated using CrysAlisPro[31)] with a fine slice method, in which the data were obtained by using the fine-slicing method with a rotation increments of $\Delta\omega = 0.01°$. Intensities of equivalent reflections were averaged and the structural parameters were refined by using JANA2006[32)].

The CDFS method[22),23)] was used to extract the entire VED distribution within the unit cell of $Ca_3Co_2O_6$ at 100 K. The [He]-type electron configuration was treated as the core-electron configuration for O, while the [Ar] configuration was used for Ca and Co. The effect of the thermal vibration was subtracted from the VED using the atomic displacement parameters determined by the high-angle analysis[22)]. Crystal structure and VED distribution were visualized by using VESTA[33)].

## 3. Results

### *3.1 Visualization of VED distribution*

Figure 2(a) shows the VED distribution of $Ca_3Co_2O_6$ at 100 K, obtained by the CDFS analysis. The local Cartesian coordinate system for each Co site is defined as $\boldsymbol{x} \parallel \boldsymbol{a}$ and $\boldsymbol{z} \parallel \boldsymbol{c}$. Yellow and orange colors represent the low- and high-level iso-density surfaces, respectively. No VED is observed around the Ca site above the displayed iso-density level, consistent with its predominantly ionic $Ca^{2+}$ character. A spherical VED distribution is observed around the O site, corresponding to the closed-shell $2s^2 2p^6$ electronic configuration of the $O^{2-}$ ion. In contrast, VED distributions with distinct anisotropies are clearly observed around the Co1 and Co2 sites, respectively. Figures 2(b) and 2(c) show the trigonal-prismatic and octahedral $CoO_6$ coordination environments, respectively, together with schematic $3d^6$ high-spin and low-spin configurations, and the calculated VED distributions around the Co1 and Co2 sites, considering only the CEF. In this model, the $3d^6$ VED $\rho_{\mathrm{Co1}}$ and $\rho_{\mathrm{Co2}}$ around the Co1 and Co2 sites are represented as

$$\rho_{\mathrm{Co1}}(\boldsymbol{r}) = |R_{\mathrm{Co}}(r)|^2 |\psi_{\mathrm{Co1}}(\theta, \phi)|^2, \quad (1)$$

$$|\psi_{\mathrm{Co1}}|^2 = |d_{3z^2-r^2}|^2 + 3/2\left|d_{x^2-y^2}\right|^2 + 3/2\left|d_{xy}\right|^2 + \left|d_{yz}\right|^2 + |d_{zx}|^2$$

and

$$\rho_{\mathrm{Co2}}(\boldsymbol{r}) = |R_{\mathrm{Co}}(r)|^2 |\psi_{\mathrm{Co2}}(\theta, \phi)|^2, \quad (2)$$

$$|\psi_{\mathrm{Co2}}|^2 = 2|d_{3z^2-r^2}|^2 + 2\left|\sqrt{2/3}d_{xy} - \sqrt{1/3}d_{zx}\right|^2 + 2\left|\sqrt{2/3}d_{x^2-y^2} - \sqrt{1/3}d_{yz}\right|^2,$$

respectively. Here, $R_{\mathrm{Co}}$ represents the radial part of the Co $3d$ orbital wavefunction, calculated using the Slater-type orbital (STO) of the isolated atom[34]. $\psi_{\mathrm{Co1}}$ ($\psi_{\mathrm{Co2}}$) describes the angular dependence of the Co1 (Co2) ion with $3d^6$ electrons. The calculated density for the low-spin configuration [Fig. 2(c)] reproduces the principal anisotropic features of the experimental VED at the Co2 site [Fig. 2(a)]. The calculation for the high-spin state [Fig. 2(b)] shows no $\phi$-dependence in the anisotropy of the calculated VED $\rho_e(\theta,\phi)$, as the minority-spin electron occupies the lowest-energy doublet $e$ orbitals with equal weight. Experimentally, however, a pronounced $\phi$-dependence is observed in the anisotropy of the observed $\rho(\theta,\phi)$ around the Co1 site [Fig. 3(a)], representing the VED at a distance $r$ = 0.2 Å from the nucleus, which corresponds to the peak of the radial profile of $\rho_{\mathrm{Co1}}(\boldsymbol{r})$ [Fig. S2[35]]. This anisotropy indicates that the equal-weight occupation of the lowest-energy $e$ doublet is insufficient and requires additional orbital mixing. This in-plane anisotropy was not observed in previous $s$-NIXS experiments[20].

### *3.2 Wavefunction analysis of the $3d^6$ high-spin state*

We analyze the observed $3d^6$ high-spin state in more detail by considering mixing between the two sets of $e$-symmetry orbitals, SOC, and $3d$-$4p$ orbital hybridization. The origin of this hybridization will be addressed later. The orbital and spin angular momentum quantum numbers for the high-spin state are $L$ = 2 and $S$ = 2, respectively. The local orbital basis functions, including the symmetry-allowed Co $4p$ admixture, are defined as

$$\psi_0 = |d_{3z^2-r^2}\rangle, \qquad (3)$$

$$\psi_{\pm1} = \frac{1}{\sqrt{2}}\left(|d_{yz}\rangle \mp i|d_{zx}\rangle\right),$$

$$\psi_{\pm2} = \frac{1}{\sqrt{2\left(1+\alpha_p^2\right)}}\left[\left(|d_{x^2-y^2}\rangle \pm i|d_{xy}\rangle\right) - \alpha_p\left(|p_x\rangle \mp i|p_y\rangle\right)\right].$$

In the absence of Co $4p$ admixture, the subscripts 0, $\pm1$, and $\pm2$ correspond to the magnetic quantum numbers of the Co $3d$ components. Based on the symmetry classification of $d$ and $p$ orbitals in $D_3$ (32) symmetry (Table 1), the $d_{x^2-y^2}$ and $d_{xy}$ orbitals can mix with the $p_x$ and $p_y$ orbitals. $\alpha_p$ is a dimensionless mixing parameter representing the relative amplitude of the Co 4$p$ component admixed into the predominantly Co 3$d$ orbital ($|\alpha_p| < 1$). We take $\alpha_p$ to be real because the spin-independent orbital mixing can be represented by a real symmetric Hamiltonian in the basis of real $d$ and $p$ orbitals. The complex combinations in Eq. (3) are subsequently constructed from these real orbitals with fixed relative phases.

We adopt the following three-component model state to describe the observed VED:

$$\psi_{\mathrm{Co1,SOC}} = A|L_z = -2, S_z = 1\rangle + B|L_z = 1, S_z = 1\rangle + C|L_z = 0, S_z = 2\rangle. \quad (4)$$

Here, $A, B$, and $C$ are wavefunction coefficients satisfying $|A|^2 + |B|^2 + |C|^2 = 1$. One should note that the phase of $C$ does not affect the VED at all. $|L_z, S_z\rangle$ denotes a spin-adapted many-electron state with total spin $S = 2$. In the pure 3$d$ limit, these states belong to the high-spin $3d^6$, $^5D$ manifold with $L = 2$. The state $|L_z, S_z = 2\rangle$ is a single Slater determinant in which each of the five orbitals contains one spin-up electron and the orbital labelled by $L_z$ additionally contains a spin-down electron. The corresponding $S_z = 1$ state is defined by $|L_z, S_z = 1\rangle = \frac{1}{2}\hat{S}_-|L_z, S_z = 2\rangle$. Here, $\hat{S}_-$ is the total-spin lowering operator. It is a normalized superposition of four Slater determinants, obtained by lowering the spin in one of the four singly occupied orbitals. Thus, the first two terms in Eq. (4) involve these spin-adapted combinations, whereas the third term involves a single Slater determinant. Angular-momentum projections are expressed in units of $\hbar$. For finite $p$ admixture, $L_z$ labels the parent 3$d$ component rather than an exact orbital-angular-momentum eigenvalue.

In ideal $D_{3h}$ ($\bar{6}m2$) symmetry, direct crystal-field mixing between the $d_{\pm 2}$ and $d_{\mp 1}$ orbital sets is forbidden. Twisting the prism [Fig. 1(b)] lowers the symmetry to 32, allowing the mixing of the $|L_z = -2, S_z = 1\rangle$ and $|L_z = 1, S_z = 1\rangle$ states. In addition, SOC further induces mixing between the $|L_z = 1, S_z = 1\rangle$ and $|L_z = 0, S_z = 2\rangle$ states. The spin-summed electron density $\rho_e$ associated with Eq. (4) is represented as

$$\rho_e = |\psi_0|^2 + 2|\psi_{\pm 1}|^2 + 2|\psi_{\pm 2}|^2 + \rho_{\mathrm{add}}, \quad (5)$$

$$|\psi_0|^2 = |d_{3z^2-r^2}|^2,$$

$$|\psi_{\pm1}|^2 = \frac{1}{2}\left(|d_{yz}|^2 + |d_{zx}|^2\right),$$

$$|\psi_{\pm2}|^2 = \frac{1}{2(1+\alpha_p^2)}\left(|d_{x^2-y^2} - \alpha_p p_x|^2 + |d_{xy} + \alpha_p p_y|^2\right),$$

$$\rho_{\text{add}} = |A\psi_{\pm2} + B\psi_{\mp1}|^2 + |C|^2|\psi_0|^2.$$

The first three terms in $\rho_e$ represent a spin-summed density contribution, not five spin-up electrons in every component of Eq. (4). The $A$ and $B$ components interfere because they have the same spin projection. Cross terms involving $C$ vanish because the density operator does not connect different spin projections. Consequently, the phase of $C$ does not affect the VED. Orthogonality of the *d* and *p* orbitals means that their overlap integrated over space vanishes; it does not require their pointwise product to vanish. Their interference can therefore contribute to the local density and, because the two components have opposite parity, can generate an inversion-asymmetric density. This model provides a basis for examining how crystal-field-induced mixing, SOC, and 4*p* admixture affect the experimentally observed VED. To capture the anisotropy of VED around the Co1 site, we optimize the coefficients $A$, $C$, and $\alpha_p$ to reproduce the anisotropy of VED $\rho(\theta,\phi)$ obtained from the CDFS analysis [Fig. 3(a)]. For fitting, we take $A$ and $C$ to be real, with $C \geq 0$ and $A^2 + C^2 \leq 1$, and set $B = \sqrt{1-|A|^2-|C|^2} \geq 0$. The $R_\rho$ value for the fitting of $\hat{\rho}(\theta,\phi)$ is defined as

$$R_\rho = \frac{\sum_{\theta,\phi}|\hat{\rho}(\theta,\phi) - s\hat{\rho}_e(\theta,\phi)|}{\sum_{\theta,\phi}|\hat{\rho}(\theta,\phi)|} \times 100. \tag{6}$$

Here, $\hat{\rho}_e(\theta,\phi) = [\rho_e(\theta,\phi) - N_e]/N_e$, $N_e$ is the number of 3*d* electrons, $\hat{\rho}(\theta,\phi) = [\rho(\theta,\phi) - \overline{\rho(\theta,\phi)}]/\overline{\rho(\theta,\phi)}$, and *s* is the scale factor. Figure 3(b) shows two-dimensional color maps of the $R_\rho$ value as a function of $A$ and $C$ at $\alpha_p$ = 0, 0.17, and 0.3. When $\alpha_p$ is fixed to 0, corresponding to no Co 4*p* orbital contribution, the lowest $R_{\rho,\min}$ value of 68.47% is found at $A = 0.32$ and $C = 0.68$ within the parameter range $-1 \leq A \leq 1$ and $0 \leq C \leq 1$. The calculated anisotropy of $\rho_e(\theta,\phi)$ [Fig. 3(c)] is similar to the experimental result [Fig. 3(a)] along the *z*-axis, but does not reproduce the *xy*-plane anisotropy. This is because a VED composed solely of even-parity *d* orbitals can sustain only the anisotropy permitted by the centrosymmetric $\bar{3}m$ site symmetry. In contrast, the experimentally observed noncentrosymmetric VED around the Co1 site with chiral site symmetry 32 necessarily involves contributions from odd-parity $p$ orbitals.

We consider the 4$p$ orbital contribution by introducing the $\alpha_p$ parameter, which results in a significant reduction of the $R_{\rho,\min}$ value to 19.98% at $\alpha_p = 0.17$, $A = 0.35$, and $C = 0.68$. The calculated $\rho_e(\theta, \phi)$ exhibits a noncentrosymmetric anisotropy with 32 site symmetry due to non-zero $\alpha_p$ [Fig. 3(d)], showing the best match with the experimental result [Fig. 3(a)]. We have reproduced the experimental VED anisotropy around high-spin $Co^{3+}$ ions. Within the present model, the fitted coefficients characterize the selected orbital-density components and their interference; they do not uniquely determine the many-electron state. Since the analysis concerns the angular density at $r = 0.2$ Å, $\alpha_p$ is treated as an effective admixture parameter rather than a direct measurement of the spatially integrated Co 4$p$ occupation.

## 4. Discussion

*4.1 Possible origin of the p orbital component around the high-spin Co site*

The experimentally observed VED anisotropy is extracted from the electron density in the immediate vicinity of the Co nucleus ($r = 0.2$ Å), where the contribution from ligand O 2$p$ orbitals is negligible, as shown in Fig. S3[35]. Therefore, the observed anisotropy cannot be attributed directly to electron density residing on the ligand atoms. Instead, it reflects the local electronic structure of the Co-centered valence orbital. At the Co1 site with 32 symmetry, direct on-site mixing between Co 3$d$ and 4$p$ orbitals is not induced by dipolar crystal-field components and can arise only through higher-order crystal-field components of octupolar symmetry. Furthermore, the large energy separation between the Co 3$d$ and 4$p$ levels is expected to strongly suppress direct on-site 3$d$–4$p$ hybridization.

In general, all Co 3$d$ orbitals hybridize with the ligand O 2$p$ orbitals, with the degree of hybridization depending on the orbital symmetry and energy; in particular, the higher-lying $e$ orbitals are expected to exhibit stronger Co 3$d$–O 2$p$ hybridization than the lower-lying $e$ orbitals. Within this general Co–O covalency, the $E$-symmetry ligand states can further hybridize with the spatially extended Co 4$p$ orbitals, providing ligand-mediated Co 3$d$–O 2$p$–Co 4$p$ pathways, as schematically illustrated in Fig. 4. Such pathways provide a possible microscopic origin of the effective Co 4$p$ component introduced in our local VED model (see Fig. S5 for details[35)]). Consequently, the experimentally reconstructed VED near the Co nucleus should not be interpreted as arising from a purely atomic Co 3*d* orbital, but rather as reflecting a hybridized local orbital containing a finite Co 4*p* admixture.

It should be emphasized that, although the CDFS analysis reconstructs the VED throughout the entire unit cell, the present discussion is based on the local VED around the Co ion, where the VED is concentrated. Therefore, the proposed Co *3d*–O 2*p*–Co 4*p* hybridization should be regarded as the most plausible microscopic interpretation of the observed local VED anisotropy, rather than as a direct experimental observation of the entire hybridized Wannier orbital. Future density functional theory calculations constrained by the experimentally determined VED may enable a more quantitative understanding of the local electronic structure, including the degree of Co 4*p* admixture.

*4.2 Wavefunction assignment of the high-spin Co site*

To reproduce the experimental VED, we assume the wave function of the $3d^6$ state as described by Eq. (4), hereafter referred to as Case 1. However, there is an alternative candidate for the wave function, expressed as

$$\psi_{\mathrm{Co1,SOC}} = A|L_z = 2, S_z = 2\rangle + B|L_z = -1, S_z = 2\rangle + C|L_z = 0, S_z = 1\rangle, \tag{7}$$

which we refer to as Case 2, where CEF mixes the $|L_z = 2, S_z = 2\rangle$ and $|L_z = -1, S_z = 2\rangle$ states, and SOC induces mixing between the $|L_z = -1, S_z = 2\rangle$ and $|L_z = 0, S_z = 1\rangle$ states. For the real coefficients and conjugate orbital pairs used here, Cases 1 and 2 give the same spin-summed VED. The VED therefore does not distinguish these two spin assignments. To assess the validity of these two wave functions, we evaluate the effective magnetic moments $\mu_{\mathrm{Co1}}$ at the high-spin Co1 site based on the obtained quantum parameters. $\mu_{\mathrm{Co1}}$ for each state is calculated as

$$\mu_{\mathrm{Co1}} = -\mu_B \langle \psi_{\mathrm{Co1,SOC}} | \hat{L}_z + 2\hat{S}_z | \psi_{\mathrm{Co1,SOC}} \rangle. \tag{8}$$

Using Eqs. (4) or (7), the effective orbital, spin, and total magnetic moments are calculated as $\langle \hat{L}_z \rangle = |A|^2(-2 + \alpha_p^2)/(1 + \alpha_p^2) + |B|^2 = 0.18$, $2\langle \hat{S}_z \rangle = 2A^2 + 2B^2 + 4C^2 = 2.92$, $\mu_{\mathrm{Co1}} = -3.11\mu_B$ for Case 1, and $\langle \hat{L}_z \rangle = |A|^2(2 - \alpha_p^2)/(1 + \alpha_p^2) - |B|^2 = -0.18$, $2\langle \hat{S}_z \rangle = 4A^2 + 4B^2 + 2C^2 = 3.08$, $\mu_{\mathrm{Co1}} = -2.89\mu_B$ for Case 2, respectively. In Case 1, the orbital magnetic moment is parallel to the spin magnetic moment, reflecting SOC that enhances the Ising anisotropy, whereas in Case 2 it is antiparallel, thereby suppressing it. Previous X-ray magnetic circular dichroism experiments reported orbital and spin magnetic moments with the same sign[18], indicating that Case 1 is more consistent with the experimental observations. The finite orbital contribution inferred for Case 1 is also consistent with previous experimental[13),18),20] and theoretical studies[10),16),19),21]. Future comparisons with density functional theory and many-body calculations, constrained by the experimentally determined VED, should enable a more quantitative understanding of the orbital moment. Nevertheless, the finite orbital contribution inferred from the present analysis provides a microscopic basis for the strong magnetocrystalline anisotropy underlying the Ising-like magnetic behavior along the *c*-axis[6)-14)].

Spectroscopic techniques provide energy-, momentum-, and polarization-resolved information on local electronic states, whereas the CDFS-based approach directly reconstructs site-resolved VED distributions in real space. It is therefore particularly sensitive to subtle anisotropies imposed by the actual crystallographic site symmetry. This real-space information complements spectroscopic and magnetic measurements and provides an independent constraint on local orbital mixing and electronic symmetry breaking.

## 5. Conclusions

We have visualized the $3d^6$ VED distributions corresponding to the low- and high-spin states of the $Co^{3+}$ ions with two different local environments in $Ca_3Co_2O_6$ using the CDFS analysis with synchrotron X-ray diffraction. The observed VED anisotropy directly reflects the spin and orbital states of the two different Co sites, providing insights into the CEF as well as the SOC and ligand-assisted Co *3d*–O 2*p*–Co 4*p* orbital hybridization. More broadly, site-resolved VED analysis provides a real-space approach for resolving spin states, orbital mixing, parity admixture, and local symmetry breaking in quantum materials containing multiple inequivalent electronic sites.

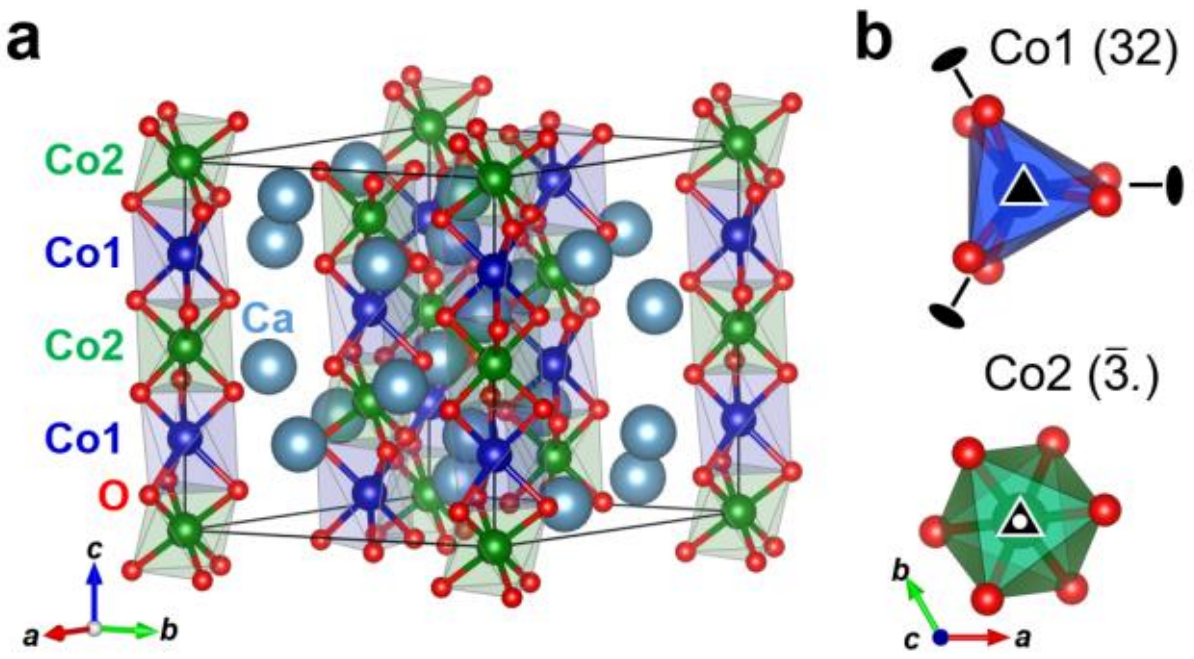

Fig. 1. (Color online) (a) Crystal structure of $Ca_3Co_2O_6$. Co1 and Co2 ions form one-dimensional chains, arranged alternately along the *c*-axis. (b) Co1 and Co2 ions, with site symmetries 32 and $\bar{3}$., are surrounded by six O atoms, forming twisted trigonal-prismatic and octahedral coordination environments, respectively. Black triangles and ellipses indicate local threefold and twofold rotation axes, respectively. A white circle on the Co2 site indicates an inversion center.

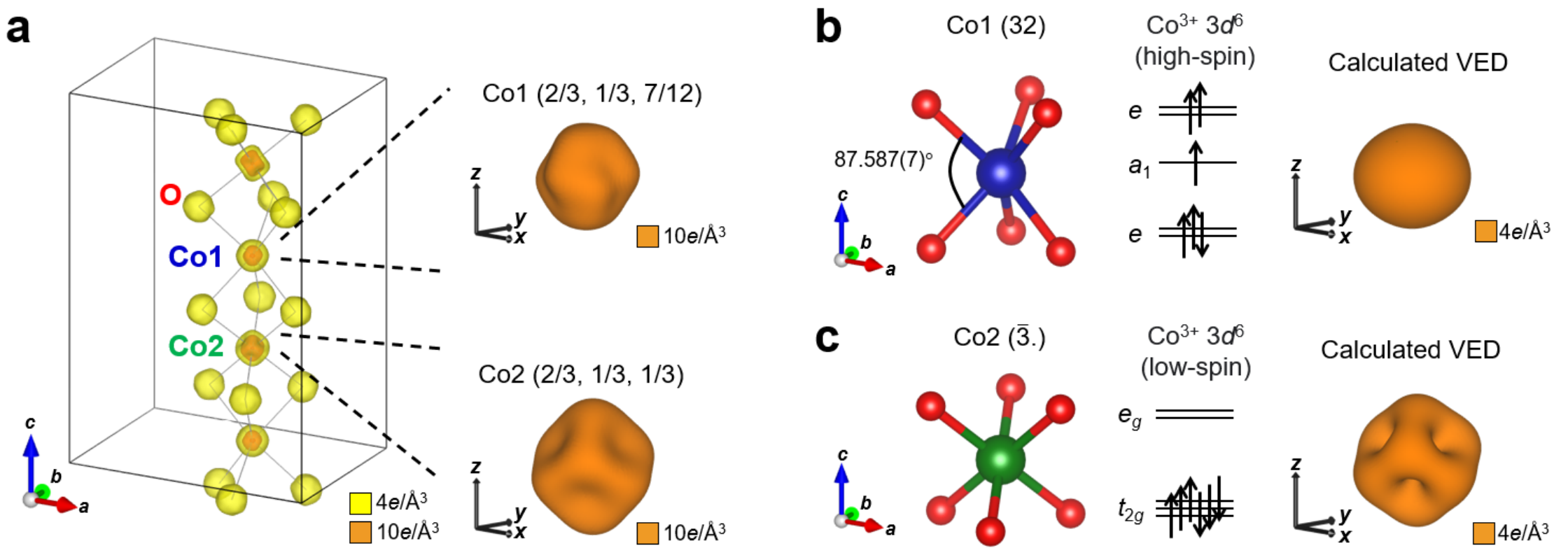

Fig. 2. (Color online) (a) VED distributions of $Ca_3Co_2O_6$ at 100 K. Yellow and orange iso-density surfaces represent different electron-density levels. (b,c) Trigonal-prismatic and octahedral coordination environments around the Co1 and Co2 sites, respectively. Schematic orbital-occupation diagrams illustrate the high-spin and low-spin $Co^{3+}$ $3d^6$ configurations. Orange iso-density surfaces show calculated VED distributions of each spin state.

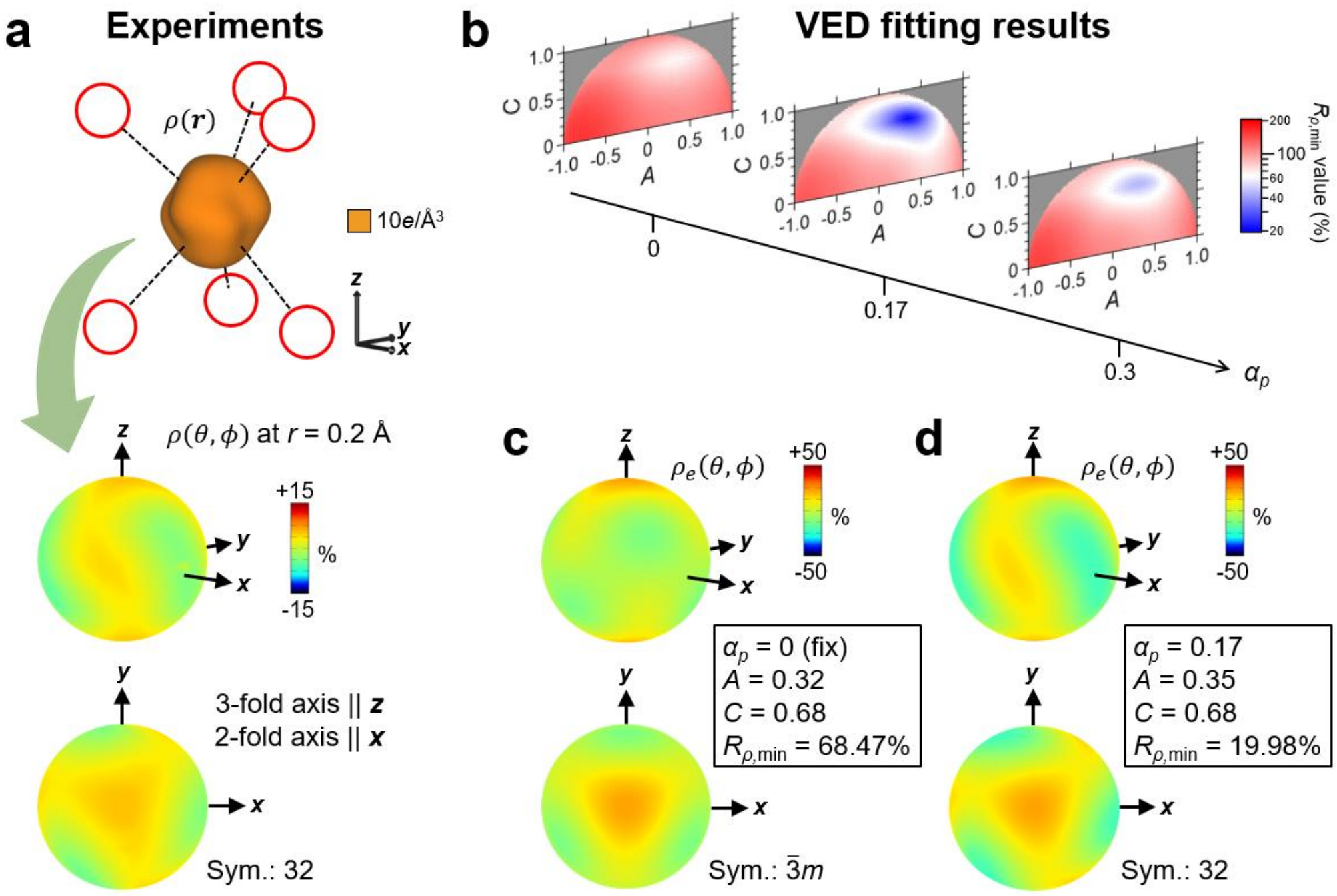

Fig. 3. (Color online) (a) VED distributions around the Co1 site in the high-spin state and its direction dependence $\rho(\theta,\phi)$ at a distance $r$ = 0.2 Å from the Co1 nucleus in the twisted trigonal-prismatic $CoO_6$ unit. The color scale on the sphere indicates $[\rho(\theta,\phi)-\overline{\rho(\theta,\phi)}]/\overline{\rho(\theta,\phi)}\times 100$ [%]. (b) Color maps of the $R_\rho$ value on two-dimensional $A$—$C$ planes at $\alpha_p = 0, 0.17, \text{and } 0.3$, with $B=\sqrt{1-|A|^2-|C|^2}$, considering the crystal-field-induced orbital mixing, SOC, and the Co 4$p$ mixing parameter in the 32 trigonal field. (c,d) Calculated angular VED distributions for the best-fit models without ($\alpha_p = 0$) and with ($\alpha_p = 0.17$) Co 4$p$ admixture, respectively. The color scale represents $[\rho_e(\theta,\phi)-N_e]/N_e\times 100$ [%], where $N_e = 6$ is the number of 3$d$ electrons.

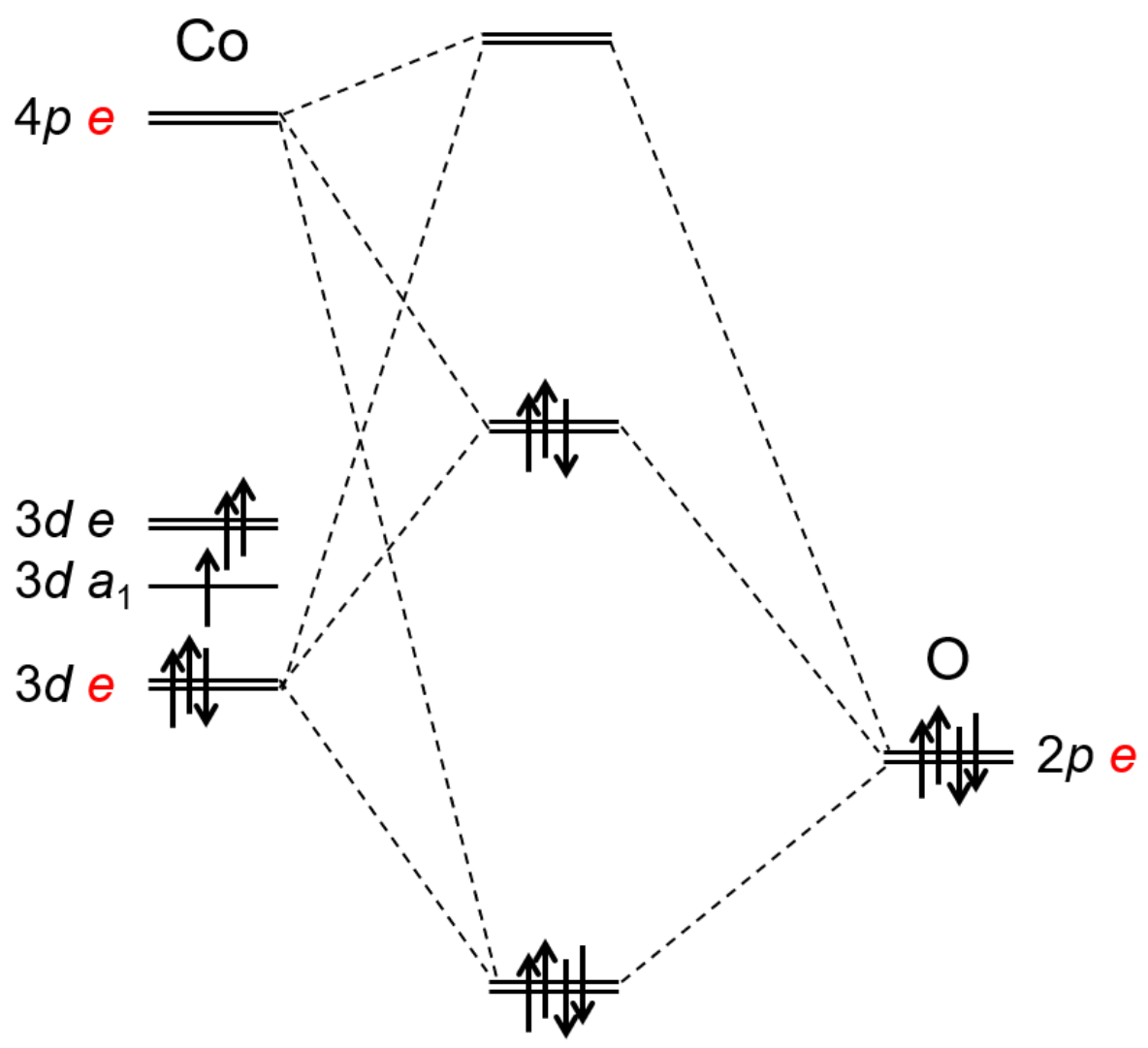

Fig. 4. (Color online) Schematic illustration of orbital hybridization among Co and O orbitals. Co 3*d*, Co 4*p*, and O 2*p* orbitals belong to the same irreducible representation *E* in $D_3$ (32) symmetry.

**Table 1.** Symmetry classification of *d* and *p* orbitals in $D_3$ (32) symmetry.

| Irrep ($D_3$) | *d* orbital | *p* orbital |
|---|---|---|
| $A_1$ | $d_{z^2}$ | |
| $A_2$ | | $p_z$ |
| $E$ | $d_{x^2-y^2}$, $d_{xy}$, $d_{xz}$, $d_{yz}$ | $p_x$, $p_y$ |

**Acknowledgment**

We thank K. Adachi and D. Hashizume for in-house X-ray diffraction characterization of the crystal quality. This work was supported by JSPS KAKENHI (Grant Nos. 22K14010, 24H01644, and 26H00626) and JST FOREST (Grant No. JPMJFR2362). The synchrotron radiation experiments were performed at SPring-8 with the approval of the Japan Synchrotron Radiation Research Institute (JASRI) (Proposal No. 2021B1261 and 2024B1599). The work at the UGC-DAE, Consortium for Scientific Research Indore was supported by the Science and Engineering Research Board (SERB), Govt. of India through Grant No. CRG/2022/005666.

†These authors contributed equally
*E-mail: kitou@edu.k.u-tokyo.ac.jp
‡E-mail: takahisa.arima@riken.jp

# Supplemental Material

Real-Space Visualization of Site-Selective Co 3*d* States in the Ising Magnet $Ca_3Co_2O_6$

Kamini Gautam[1], Shunsuke Kitou[1,2], Yuiga Nakamura[3], Norimasa Sasabe[4], Arvind Kumar Yogi[5], Dinesh Kumar Shukla[5], Yuichi Yamasaki[1,4,6], and Taka-hisa Arima[1,2]

[1]*RIKEN Center for Emergent Matter Science (CEMS), Wako 351-0198, Japan*

[2]*Department of Advanced Materials Science, The University of Tokyo, Kashiwa 277-8561, Japan*

[3]*Japan Synchrotron Radiation Research Institute (JASRI), SPring-8, Hyogo 679-5198, Japan*

[4]*Center for Basic Research on Materials (CBRM), National Institute for Materials Science (NIMS), Tsukuba, 305-0047, Japan*

[5]*UGC-DAE Consortium for Scientific Research, Khandwa Road, Indore 452001, India*

[6]*International Center for Synchrotron Radiation Innovation Smart, Tohoku University, Sendai 980-8577, Japan*

## Crystal structure and VED distribution of $Ca_3Co_2O_6$ at 100 K

The results of the structural analysis of $Ca_3Co_2O_6$ at 100 K are summarized in Fig. S1 and Tables S1-S2. CCDC 2566488 contains the supplementary crystallographic data for this paper. These data can be obtained free of charge via www.ccdc.cam.ac.uk/data_request/cif.

Figure S1(a) shows the logarithmic $|F_O|^2$-$|F_C|^2$ plot as a result of the structural analysis. $F_O$ and $F_C$ correspond to the observed and calculated crystal structural factors, respectively. Figure S1(b) shows part of the crystal structure of $Ca_3Co_2O_6$ at 100 K with atoms represented with 99% probability displacement ellipsoids. Figure S1(c) shows a Co1$O_6$ trigonal-prism that is twisted by 14.6°.

The red and blue solid lines in Fig. S2 represent radial profiles of the valence electron density (VED) around the trigonal-prism Co1 sites in the $-a$ and $+c$ directions, respectively. The gray dashed line in Fig. S2 represents the electron density of the Co $3d^6$ orbital, calculated using the Slater-type orbital (STO) of an isolated atom [1]. The black solid line in Fig. S2 represents the electron density at the same resolution as the experiment ($d_{min}$ = 0.28 Å), calculated following the procedures described in Ref. [2].

Figure S3 shows one-dimensional profiles of the electron densities of the Co-3$d$ and O-2$p$ orbitals, calculated using the STO [1]. While the Co-3$d$ electron density is plotted from the origin at $r = 0$ Å in the positive direction, the O-2$p$ electron density is plotted from $r = 2.06578$ Å in the negative direction. This distance corresponds to the experimentally determined Co1-O bond length in the Co1$O_6$ trigonal-prism. In Fig. 3(b) of the main manuscript, the anisotropy of the observed VED $\rho(\theta,\phi)$ around the Co1 site is plotted at $r = 0.2$ Å from the Co1 nucleus, indicating that the contribution from the O-2$p$ orbitals to this anisotropy is negligible.

## Hybridizations among Co 3*d*, Co 4*p*, and O 2*p* orbitals

Since the Co $3d(e)$, O $2p_L(e)$, and Co $4p(e)$ states belong to the same irreducible representation $E$ in the $D_3$ (32) point group, they can hybridize within the same Hilbert space. A schematic illustration of the ligand molecular orbitals (LMOs) for the ideal $D_{3h}$ ($\bar{6}m2$) symmetry is shown in Figure S4. Upon lowering the symmetry from $\bar{6}m2$ to 32 through twisting of the trigonal prism, the $E'$ and $E''$ LMOs both transform as the $E$ irreducible representation, allowing them to hybridize and form the $E$-symmetry LMOs considered in the present analysis. In the basis $\{|3d(e)\rangle, |2p_L(e)\rangle, |4p(e)\rangle\}$, the effective Hamiltonian is written as

$$H_{e'} = \begin{pmatrix} \varepsilon_d & t_{dL} & t_{dp} \\ t_{dL}^* & \varepsilon_L & t_{Lp} \\ t_{dp}^* & t_{Lp}^* & \varepsilon_p \end{pmatrix},$$

where $\varepsilon_d$, $\varepsilon_L$, and $\varepsilon_p$denote the on-site energies of the Co $3d(e)$, O $2p_L(e)$, and Co $4p(e)$states, respectively. The off-diagonal terms $t_{dL} = \langle L \left| \frac{p^2}{2m} + V(r) \right| 3d \rangle$, $t_{Lp} = \langle 4p \left| \frac{p^2}{2m} + V(r) \right| L \rangle$, and $t_{dp} = \langle 4p|V(r)|3d\rangle$ describe the $3d-2p_L$, $2p_L-4p$, and direct $3d-4p$ hybridizations. The hybridized eigenstate is expressed as

$$|e'\rangle = a\,|3d(e)\rangle + b\,|2p_L(e)\rangle + c\,|4p(e)\rangle.$$

To clarify the origin of the $4p$ admixture, we consider the limit $\varepsilon_d \approx \varepsilon_L$, while $\varepsilon_p$ is significantly higher in energy. In this regime, the dominant hybridization occurs between the Co $3d$ and ligand $2p$ orbitals. Neglecting the direct $3d–4p$ hopping in the first step, the $3d$–ligand subspace is diagonalized as

$$H_{dL} = \begin{pmatrix} \varepsilon_d & t_{dL} \\ t_{dL}^* & \varepsilon_L \end{pmatrix},$$

yielding $|\psi_\pm\rangle = a_\pm|d\rangle + b_\pm|L\rangle$ with eigenenergies

$$E_\pm = \frac{\varepsilon_d + \varepsilon_L}{2} \pm \sqrt{\left(\frac{\varepsilon_d - \varepsilon_L}{2}\right)^2 + |t_{dL}|^2},$$

where $|\psi_+\rangle$ and $|\psi_-\rangle$ represent the antibonding and bonding molecular orbitals, respectively. For $\varepsilon_d \approx \varepsilon_L$, these reduce to $|\psi_\pm\rangle \simeq (|d\rangle \pm |L\rangle)/\sqrt{2}$.

Since the Co $4p$ level lies at a much higher energy (Figure S5), the low-energy electronic structure is governed primarily by these $3d$–ligand molecular orbitals. Among them, the antibonding state $|\psi_+\rangle$ is particularly important, as it lies closer in energy to the Co $4p$ level and thus provides the dominant channel for additional hybridization. In the basis $\{|\psi_-\rangle, |\psi_+\rangle, |4p\rangle\}$, the Hamiltonian becomes

$$H_{e'} = \begin{pmatrix} E_- & 0 & t_{-p} \\ 0 & E_+ & t_{+p} \\ t_{-p}^* & t_{+p}^* & \varepsilon_p \end{pmatrix},$$

where $t_{\pm p} = \langle \psi_\pm | H_{e'} | 4p\rangle = a_\pm t_{dp} + b_\pm t_{Lp}$. Because the direct $3d–4p$ coupling is weak, we may approximate $t_{\pm p} \simeq b_\pm t_{Lp}$, and for $\varepsilon_d \approx \varepsilon_L$, $t_{\pm p} \simeq t_{Lp}/\sqrt{2}$. Because $E_+ > E_-$, the antibonding state lies closer to the $4p$ level, i.e., $\varepsilon_p - E_+ < \varepsilon_p - E_-$, and therefore acquires a larger $4p$ admixture. To estimate this admixture explicitly, we consider the subspace spanned by $\{|\psi_+\rangle, |4p\rangle\}$, for which the Hamiltonian is

$$H_{+p} = \begin{pmatrix} E_+ & t_{+p} \\ t_{+p}^* & \varepsilon_p \end{pmatrix}.$$

We now seek the lower-energy eigenstate in the form $|\Phi_+\rangle = A|\psi_+\rangle + C|4p\rangle$. The Schrödinger equation $H_{+p}|\Phi_+\rangle = E|\Phi_+\rangle$ gives

$$(E_+ - E)A + t_{+p}C = 0,$$
$$t_{+p}^* A + (\varepsilon_p - E)C = 0.$$

From the second equation, the ratio of the $4p$ component to the $|\psi_+\rangle$ component is

$$\frac{C}{A} = -\frac{t_{+p}^{*}}{\varepsilon_p - E}.$$

When the hybridization is weak compared with the level separation, the lower eigenvalue is approximately $E \cong E_{+}$, and thus

$$\frac{C}{A} \cong -\frac{t_{+p}^{*}}{\varepsilon_p - E_{+}}.$$

Therefore, the magnitude of the $4p$ admixture in the antibonding state is estimated as

$$\left|\frac{C}{A}\right| \cong \frac{|t_{+p}|}{\varepsilon_p - E_{+}}.$$

Using $t_{+p} \simeq b_{+} t_{Lp}$ and $b_{+} \simeq 1/\sqrt{2}$, one obtains

$$\left|\frac{C}{A}\right| \cong \frac{|t_{Lp}|}{\sqrt{2}(\varepsilon_p - E_{+})}.$$

Furthermore, for $\varepsilon_d \simeq \varepsilon_L$,

$$E_{+} \simeq \varepsilon_d + |t_{dL}|,$$

so that

$$\left|\frac{C}{A}\right| \cong \frac{|t_{Lp}|}{\sqrt{2}(\varepsilon_p - \varepsilon_d - |t_{dL}|)}.$$

This analysis demonstrates that the relevant low-energy electronic state of $e$ symmetry is not a pure Co $3d$ state but an antibonding $3d$–ligand Wannier orbital with an admixture of Co $4p$ character. Using representative parameters of $t_{dL} \cong 1.5$ eV, $t_{Lp} \cong 1.4$ eV, and $\varepsilon_p - \varepsilon_d \cong 6.45$ eV, the estimated ratio $C/A$ is approximately $0.2$, consistent with the experimentally determined $4p$ admixture ($\alpha_p = 0.17$) reported in the main text.

Table S1. Summary of crystallographic data of $Ca_3Co_2O_6$ at 100 K.

| | |
|---|---|
| Wavelength (Å) | 0.3093 |
| Crystal dimension ($\mu m^3$) | 65 × 40 × 30 |
| Space group | $R\bar{3}c$ |
| $a$ (Å) | 9.07830(10) |
| $c$ (Å) | 10.38560(10) |
| Z | 6 |
| $F$(000) | 972 |
| $(\sin\theta/\lambda)_{max}$ (Å$^{-1}$) | 1.79 |
| $N_{total}$ | 48737 |
| Average redundancy | 12.564 |
| Completeness (%) | 97.19 |
| Number of unique reflections ($I$>3$\sigma$ / all) | 3603 / 3879 |
| $N_{parameters}$ | 19 |
| $R_1$ ($I$>3$\sigma$ / all) (%) | 1.77 / 1.86 |
| $wR_2$ ($I$>3$\sigma$ / all) (%) | 2.73 / 2.75 |
| GOF ($I$>3$\sigma$ / all) | 1.86 / 1.80 |

Table S2. Structural parameters of $Ca_3Co_2O_6$ at 100 K.

| Atom | Wyckoff position | Site symmetry | $x$ | $y$ | $z$ |
|---|---|---|---|---|---|
| Co(1) | 6*a* | 32 | 1/3 | 2/3 | 5/12 |
| Co(2) | 6*b* | $\bar{3}$. | 1/3 | 2/3 | 1/6 |
| Ca | 18*e* | .2 | 1/3 | 0.297369(5) | 5/12 |
| O | 36*f* | 1 | 0.308959(18) | 0.490224(18) | 0.280252(13) |

| Atom | $U_{11}$ ($10^{-3}$ Å$^2$) | $U_{22}$ ($10^{-3}$ Å$^2$) | $U_{33}$ ($10^{-3}$ Å$^2$) | $U_{12}$ ($10^{-3}$ Å$^2$) | $U_{13}$ ($10^{-3}$ Å$^2$) | $U_{23}$ ($10^{-3}$ Å$^2$) |
|---|---|---|---|---|---|---|
| Co(1) | 3.682(10) | = $U_{11}$ | 2.593(14) | 1.841(5) | 0 | 0 |
| Co(2) | 1.800(9) | = $U_{11}$ | 1.808(12) | 0.900(4) | 0 | 0 |
| Ca | 3.163(12) | 3.000(10) | 3.278(13) | 1.582(6) | 0.364(7) | 0.182(3) |
| O | 4.84(3) | 3.69(3) | 4.22(3) | 2.31(2) | 0.07(2) | 0.81(2) |

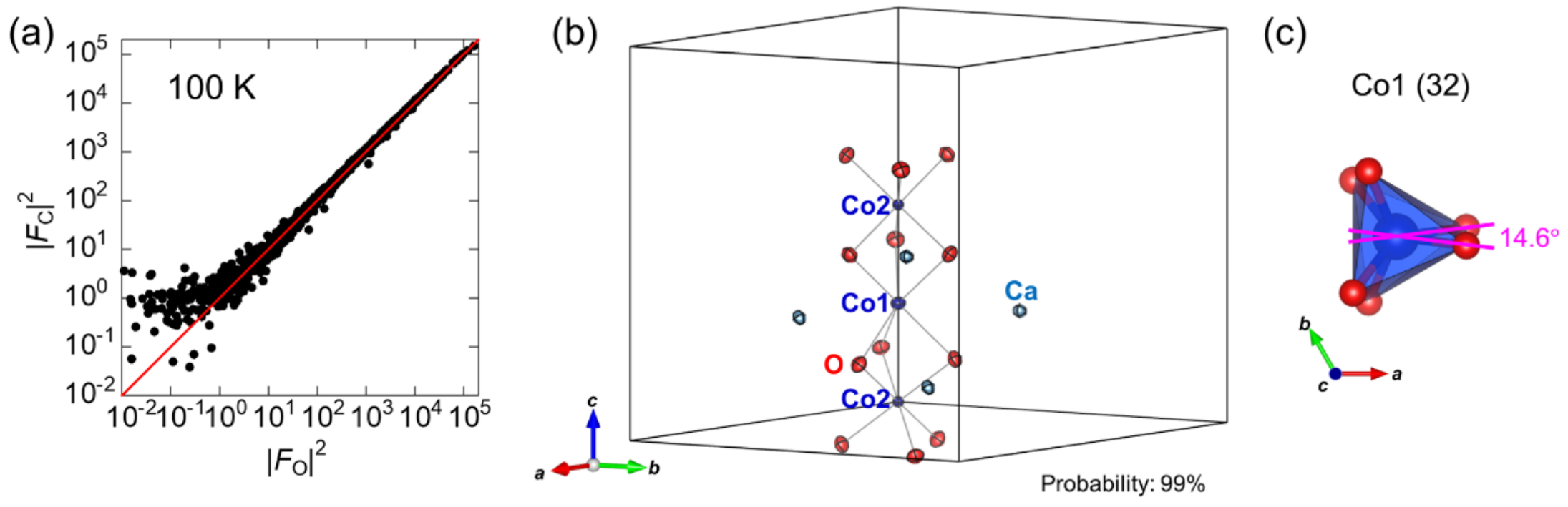

FIG. S1. (a) Logarithmic $|F_O|^2$ versus $|F_C|^2$ plot as a result of the structural analysis of $Ca_3Co_2O_6$ at 100 K. (b) Crystal structure of $Ca_3Co_2O_6$ at 100 K, with atoms represented by displacement ellipsoids (99% probability).

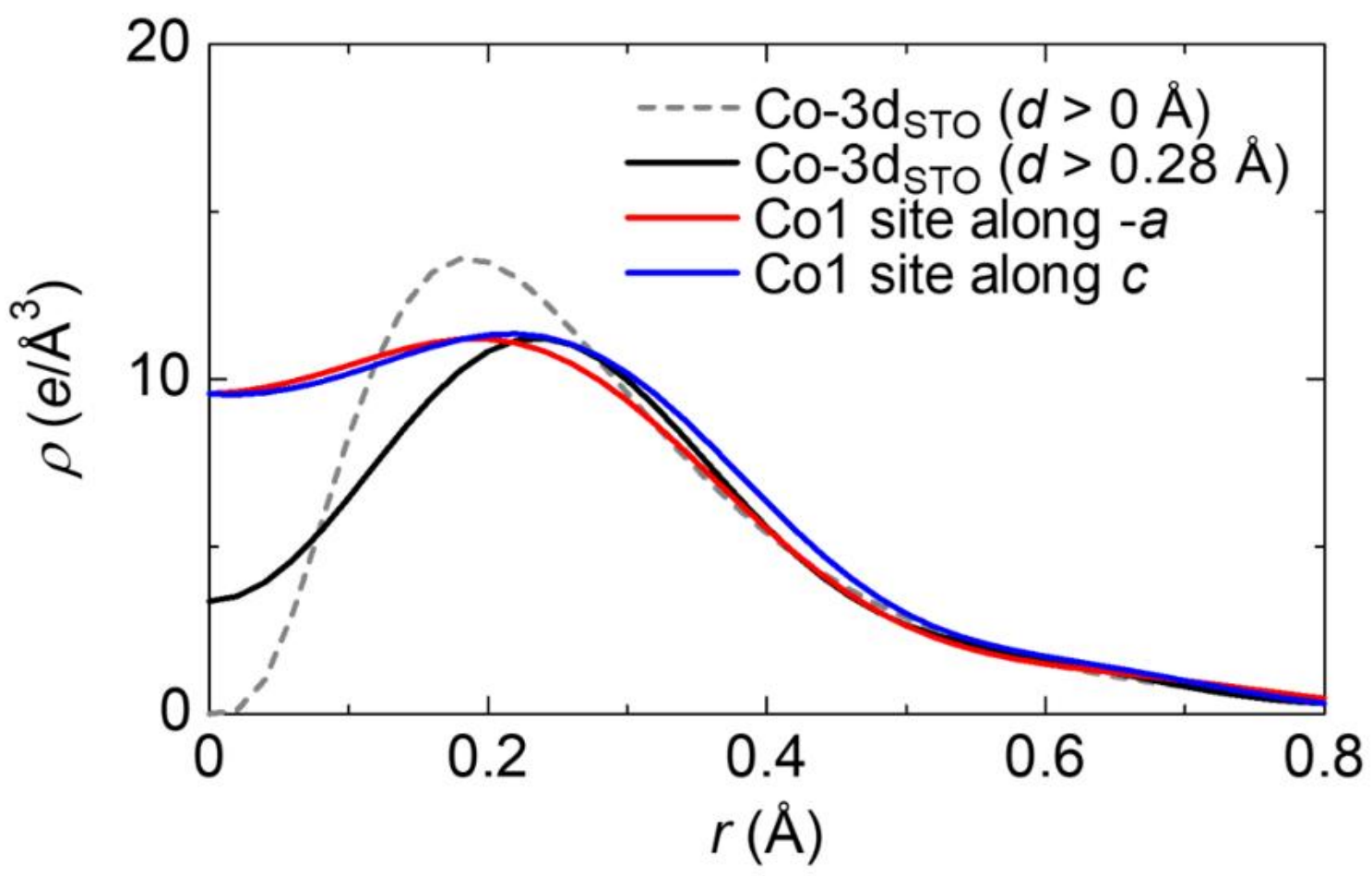

FIG. S2. Radial profiles of the calculated (gray dashed and black solid lines) and experimentally observed (red and blue solid lines) VEDs around the Co atom.

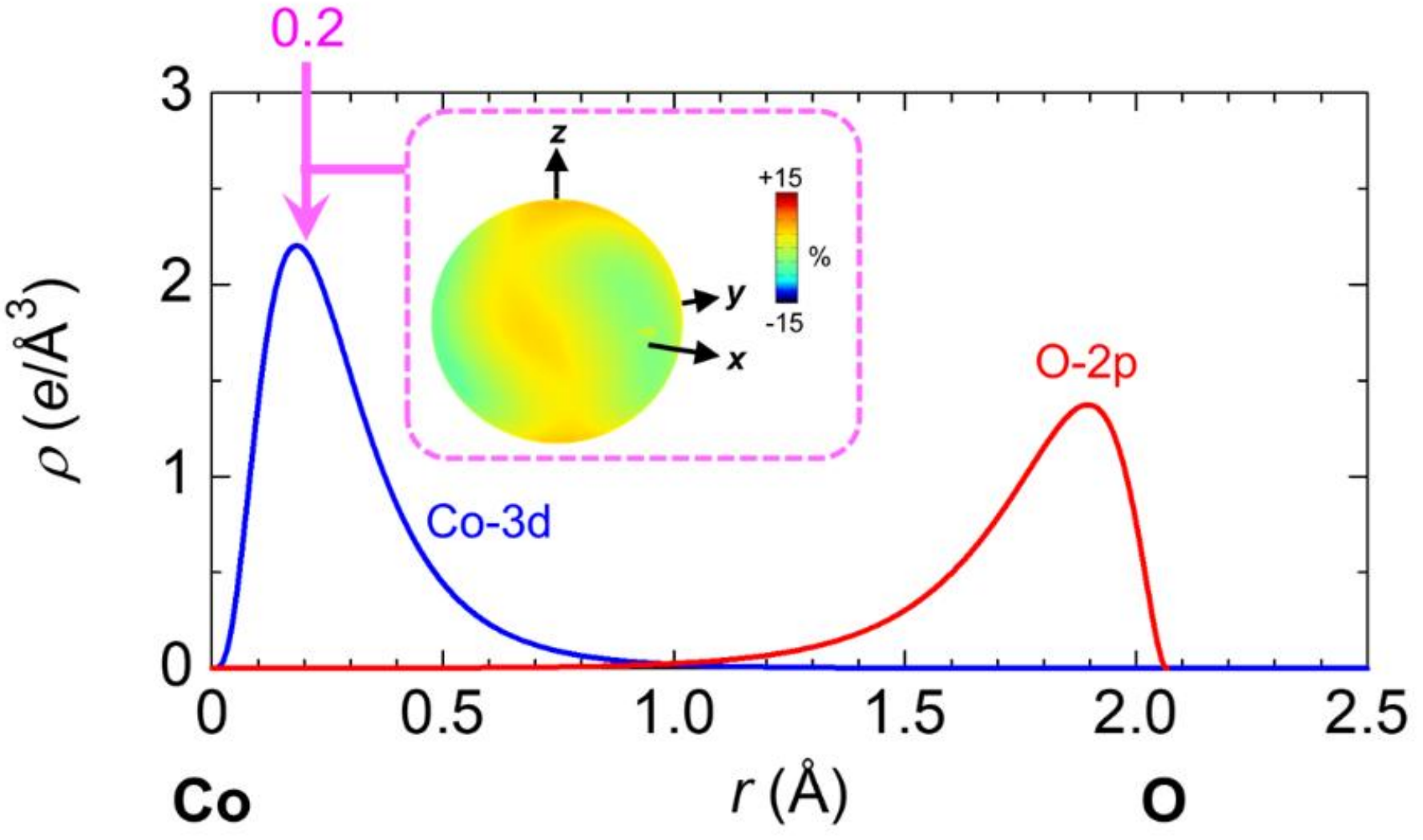

FIG. S3. Radial profiles of the electron densities of the Co-3$d$ (blue line) and O-2$p$ (red line) orbitals. The inset enclosed by pink dotted lines shows the direction dependence of the observed VED $\rho(\theta,\phi)$ at a distance $r$ = 0.2 Å from the Co1 site.

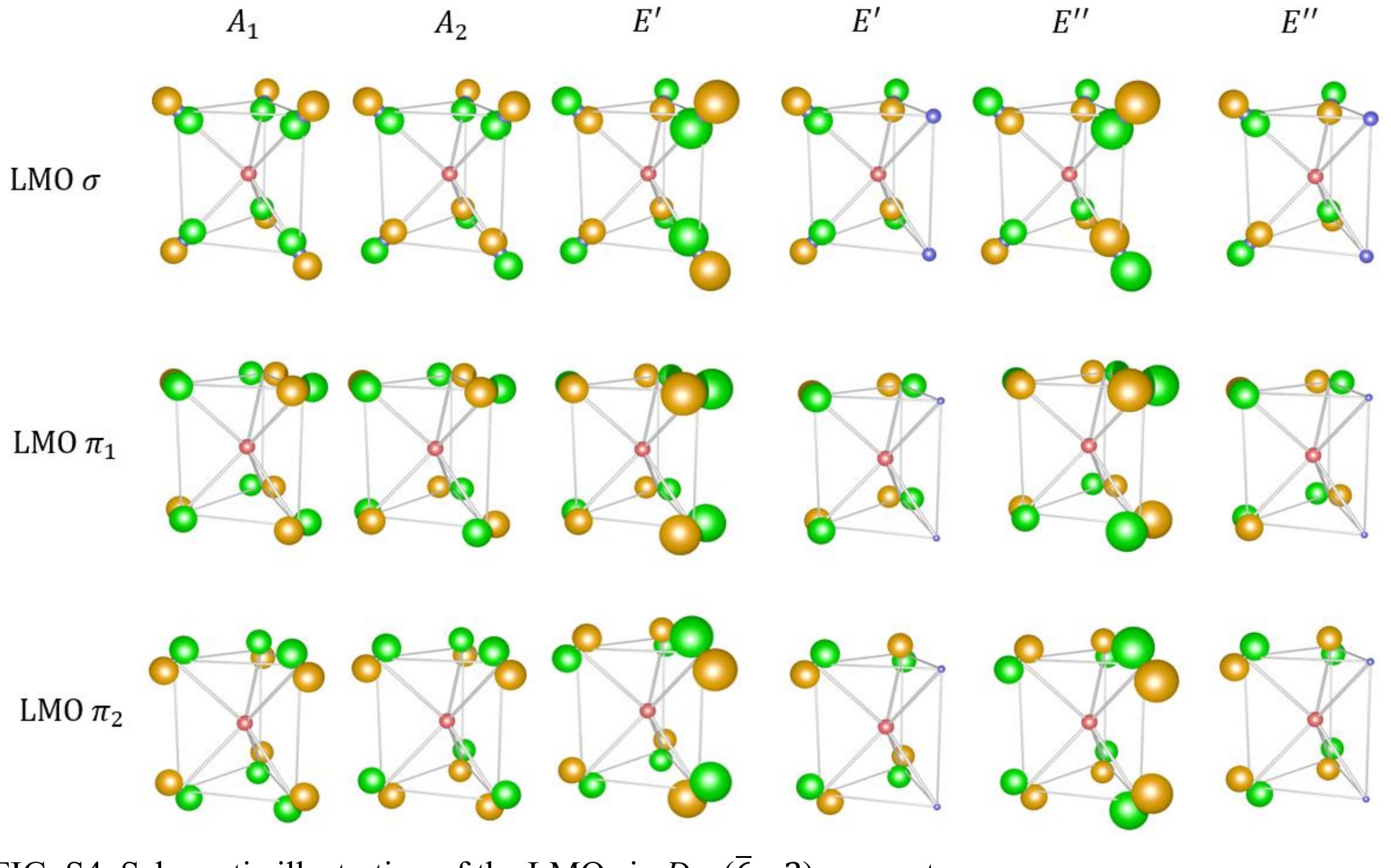

FIG. S4. Schematic illustration of the LMOs in $D_{3h}$ ($\bar{6}m2$) symmetry.

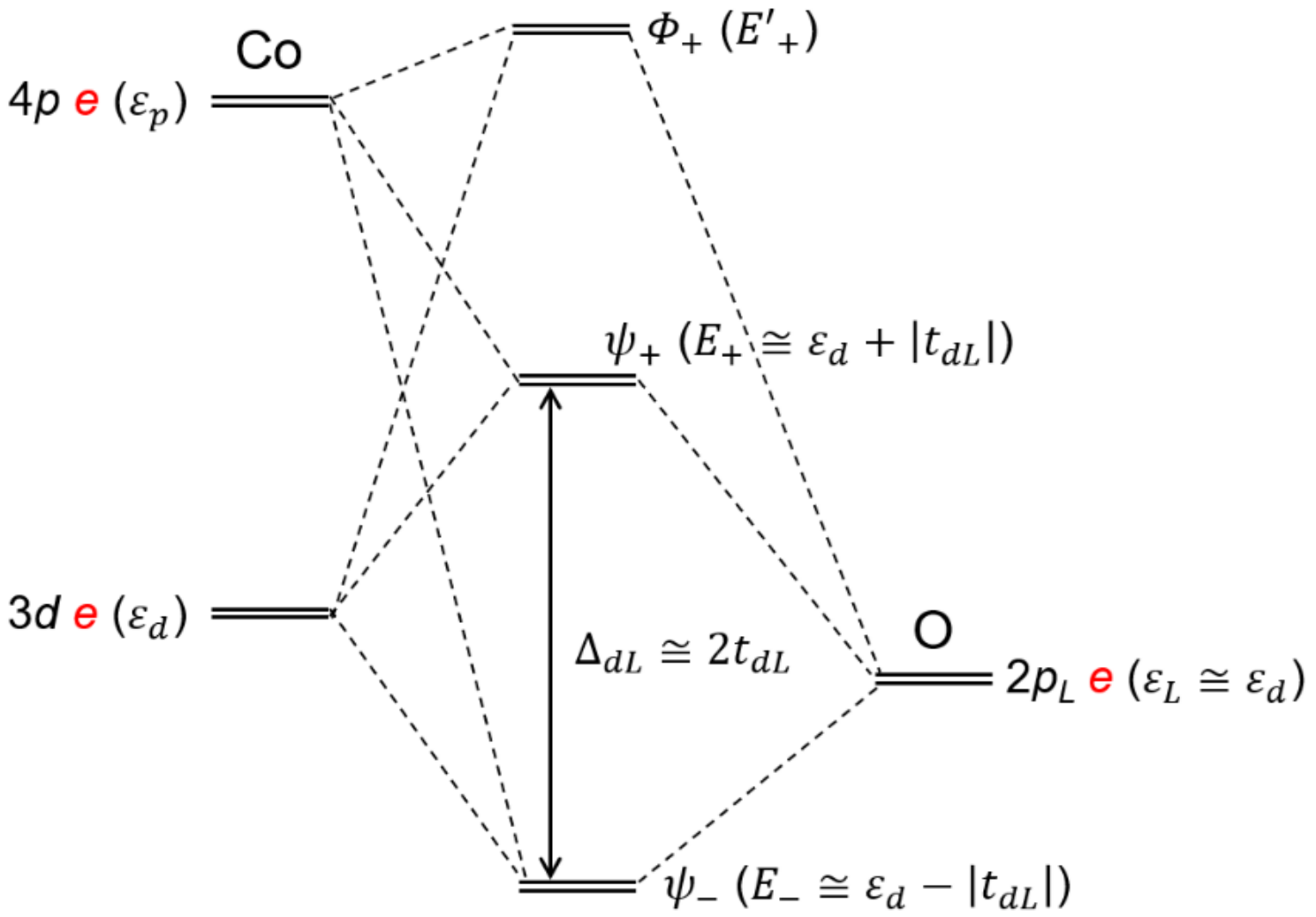

FIG. S5. Schematic illustration of hybridization among Co $3d$, Co $4p$, and O $2p_L$ orbitals, which belong to the same irreducible representation $E$ in $D_3$ (32) symmetry.